# Orphan risks at the frontier of artificial intelligence: What diverging safety and compliance frameworks reveal about how AI companies choose the risks they prioritize

Andrew D. Maynard

*School for the Future of Innovation in Society, Arizona State University, Tempe, AZ, USA. e-mail: andrew.maynard@asu.edu*

**Abstract**

Companies developing some of the world's most powerful artificial intelligence systems are surprisingly diligent in how they map out the risks their technologies present. Yet the risk landscape that lies between emerging frontier models and their economically successful and societally beneficial deployment is becoming increasingly hard to navigate. Complicating this further, many frontier AI companies maintain more than one account of what could go wrong with their technologies. This paper documents the divergence between these accounts by comparing safety and compliance documents published by Anthropic, OpenAI, Google DeepMind and Meta between 2023 and 2026, and considers what the resulting record reveals about how these companies select the risks they manage. As these documents are timestamped and archived, they provide a valuable public record of institutional risk selection in progress. From this record the paper identifies four filters that determine which risks tend to survive in self-authored frameworks (measurability, severity, auditability and competitive cost) and introduces the "safety differential" as the gap between the risk landscape a company selects for itself, and the one regulators select for it. While acute, quantifiable risks appear across documents, less tractable risks such as harmful manipulation are articulated fluently where law compels disclosure, yet remain absent from most self-chosen frameworks. This is an exclusion that follows from how these institutions define risk. Drawing on scholarship on institutional risk selection and the framework of risk innovation, the paper shows how redefining risk as a threat to value can help explain how risks become "orphan risks," how it indicates where future blindsides may occur, and how it points to lightweight tools for de-orphaning risks that frontier AI's safety apparatuses are not currently organized to address.

**Keywords:** artificial intelligence, frontier AI, risk, safety, AI risk, AI safety, risk innovation, orphan risk, severity floor, safety differential

## 1. Introduction

In December 2023, OpenAI published the first version of its Preparedness Framework — a document in which the company publicly set out potential risks associated with its most capable models (a class of systems now called "frontier" AI), and one where the company made a commitment to track and manage these.[1] Four key categories of risk made the list: cybersecurity; chemical, biological, radiological and nuclear threats; model autonomy; and persuasion — which the framework described in terms of models being used to convince people to change their beliefs. In OpenAI's framework, persuasion was treated with the same degree of seriousness as the other three risk categories. It had its own graded scale, from low to critical, and its own place in the machinery of evaluations and thresholds that the framework built around each tracked risk.

Sixteen months later, persuasion was gone. In effect, potential risks associated with persuasion were "orphaned," suggesting an emerging landscape around frontier AI models where some risks are attended to more than others, and where the level of attention a risk receives may not necessarily align with its societal importance. It's this possibility that this paper explores, and applies the framework of risk innovation to as a potential route to addressing orphaned but nevertheless important AI risks.

In OpenAI's case, the orphaning of persuasion as a risk came with the second version of their framework, published in April 2025. This version removed persuasion from the tracked categories, explaining that risks of this kind do not fit the criteria for a framework aimed at preventing specified severe harms. These were harms that the revision now defined explicitly as "the death or grave injury of thousands of people or hundreds of billions of dollars of economic damage".[2] According to OpenAI, persuasion-related risks would instead be handled through the company's usage policies and its investigations of misuse. In other words, the company's internal responses to persuasion risks were real, but were discretionary, with no published thresholds, no pre-committed responses and — importantly — no "versioned" public record of their evolving thinking, commitments, and actions. The upshot was that a risk that had spent a year and a half on the public commitment list moved into a less-public discretionary layer of assessment and management.

Then, in May 2026, it came back. This was when OpenAI published its Frontier Governance Framework — a document that was produced in response to new legal requirements; specifically, California's Transparency in Frontier Artificial Intelligence Act — which requires large frontier developers to publish frameworks addressing catastrophic risk — and the binding obligations that the European Union's AI Act places on providers of general-purpose models.[3,4,5] OpenAI, like most of its major rivals, committed to meeting those European obligations by signing onto the EU's General-Purpose AI Code of Practice.[6] The code is admittedly voluntary — companies are not compelled to sign — but the obligations behind it are driven by law, which makes signing it a pragmatic path of least resistance for AI developers. Among the risk categories the May 2026 document covers is harmful manipulation: the strategic distortion of what people believe and do at scale, including influence operations and election interference. This, in effect, covers much of the same territory that the persuasion category had covered, but under a new and somewhat narrower name.

OpenAI is candid in noting that its approach to this category is exploratory and less developed than its work on other areas of risk. And yet the risk is named, owned and reported on. In other words, the category the company had judged unsuited to its own framework had returned. Despite this, there is little evidence that there was a sudden increase in the probability of persuasion- or manipulation-based risk in the intervening year (although as frontier models develop, the risk landscape is constantly evolving). Rather, what had changed was that regulators in California and the EU, and not the company, were beginning to influence the list of publicly identified risk categories.

Here, OpenAI's story is just one example of a pattern that can be seen running across the industry. Anthropic, the frontier model developer founded by former OpenAI researchers, has never tracked persuasion explicitly in its own safety framework. This, to be clear, is not because the company ignores such risks: the system cards it publishes with each model discuss associated concerns ranging from sycophancy to user wellbeing in some depth. But these cards describe what a model *does*. In contrast, the safety frameworks define what the company has committed to *manage*. And at the framework level, a 2024 revision of Anthropic's Responsible Scaling Policy set persuasion aside as "not yet sufficiently understood to include in our current commitments".[7] Yet when Anthropic published its Frontier Compliance Framework in December 2025 to meet the Californian statute (and, over time, its European obligations), the pull of the broader list of risks that regulators care about began to show, and within months (with European enforcement approaching) a revision had added a risk tier on harmful manipulation.[8,9]

These examples from OpenAI and Anthropic reflect practices within private companies. But there is also a telling signal here coming from public companies. Meta and Alphabet — two publicly traded companies among the four developers this paper focuses on — are legally required to describe risks that are material to their business in their annual securities filings. Here, Meta names misinformation, harmful content and youth safety among them: risks that its own safety framework implicitly places out of scope. Alphabet makes a similar move, devoting a dedicated risk factor to the reputational harm that AI failures could inflict on its business.[10,11]

What emerges is companies assessing the same models, but describing different risk landscapes in documents written for different audiences. And as a consequence, which landscape comes into view depends on who is doing the asking. Interrogate a company itself through self-authored safety frameworks, and the landscape tends to feature a handful of catastrophic capabilities. But interrogate it through a statute or a code, and a broader landscape emerges — and one that includes risks like manipulation. Take this one step further (as in the case of Meta and Alphabet) and interrogate through securities law, and risks like misinformation, youth safety and reputational damage appear.

Of course, the source documents here serve different purposes, and setting them side by side is not a like-for-like comparison. Yet even allowing for that, the comparison reveals where law compels articulation, risks are articulated fluently — even complex risks like manipulation. In other words, the companies are able to describe these risks, even though their internal safety documents tend to skirt around them. What keeps them out of these self-authored frameworks, I will argue, is the ways those frameworks define and select risk.

The arguments I develop below lean on the differences between these documents. And because of this it's worth being clear about what each type of document represents — not least because the boundaries are perhaps blurrier than the labels might suggest. The *safety frameworks* are the documents that companies write for themselves: Anthropic's Responsible Scaling Policy, OpenAI's Preparedness Framework, Google DeepMind's Frontier Safety Framework, and Meta's Advanced AI Scaling Framework.[2,7,12,13] These are not required by law. Rather, they are documents that map out the risks which a company decides it will be publicly accountable for. And in most cases they establish accountability around thresholds, evaluations and committed responses. In comparison, the *compliance frameworks* are the documents that emerging AI-focused laws require. California's Transparency in Frontier Artificial Intelligence Act demands them outright. In contrast, Europe's governance lever is the General-Purpose AI Code of Practice: voluntary to sign, but binding in what it requires once the AI Act stands behind it.[3,4,5,6] *Securities filings* on the other hand are enumerations of whatever could materially harm a business, and that are legally required to be disclosed to investors. Alongside all of these sit the *system cards* that now accompany each major model release. These are reports on how a particular model performed in evaluations and assessments against whatever its developer's framework tracks. System cards can range widely, and can describe in detail what a model does and what its capabilities and vulnerabilities are in ways that safety frameworks don't necessarily track. But while they may describe model behavior and implications in detail, they rarely come with commitments to action — which is why they are peripheral to the analysis here.

In what follows I concentrate on four developers — Anthropic, OpenAI, Google DeepMind and Meta — because their frameworks are perhaps the most consequential and the best documented. But of course they are not the only players here. Other developers in the United States and beyond publish frameworks of their own. And Chinese companies in particular operate under an increasingly elaborate state-guided regime. At some point a comparative study across these would be valuable. But this is beyond the scope of this paper. What is within scope is how different safety and risk-related documents from the four companies identified above reveal strengths and weaknesses around how risk landscapes are developed and used, and how this might inform alternative approaches to understanding and navigating frontier AI system risks. Within this scope, it's worth noting that safety frameworks are no mere formality. Pioneered by individual developers in 2023 and generalized when sixteen organizations signed the Frontier AI Safety Commitments at the 2024 AI Seoul Summit, these documents have become the de facto governance layer for frontier AI — the place where decisions about when to develop, when to deploy, and when to stop, are given public form.[14] Governments have increasingly chosen to build on these frameworks rather than simply replace them. California, for instance, now requires large developers to publish such frameworks, and Europe's new obligations lean heavily on a code of practice that intersects with the frameworks.[4,6] Which risks these documents focus on, and which they leave out, is therefore not simply an internal matter. Rather, what these documents cover and what they do not has become indicative of who decides what counts as an AI risk worth managing, and in what context.

Here, I would argue that the emerging pattern of managed and unmanaged risk in frontier AI as indicated in these documents is neither accidental nor, for the most part, cynical. Rather, it is the result of a risk-selection process that follows from how these frameworks (and the people and organizations who develop them) define risk. And, unusually for such processes, it is one that has left a public trail that can be studied. Published frameworks are versioned, timestamped and archived. In effect their revisions can be read as a way to gain insights into how AI companies are mapping and responding to potential risks within multiple contexts.

In what follows I consider that record, identify four filters that determine which risks survive in what is published, trace three of those filters to definitions of risk inferred to be in use and the fourth to the competitive environment, and then ask what changes if risk is defined differently — in this case, drawing on the definition used in the risk innovation framework of risk as a threat to what people, organizations and societies value. This redefinition, I will argue, can help explain the patterns observed, and anticipate where emerging risks may lead to unexpected harm. The redefinition also comes with working tools — again drawing on the risk innovation framework — that have the potential to, in effect, "de-orphan" orphaned risks that have the ability to lead to harm in ways that are not conventionally anticipated.

## 2. Blank spaces on a crowded map

If managing the risks of frontier AI systems came down to simply identifying the risks, navigating the emerging risk landscape would be much easier than it is. But of course this is not how risk management works. I mention this as AI risk-naming has already been done at a remarkable scale. The MIT AI Risk Repository, for instance — a living synthesis of published taxonomies — now covers more than 1,600 distinct potential risks, from discrimination and misinformation through to catastrophic misuse.[15] And the International AI Safety Report, a government-mandated scientific assessment chaired by Yoshua Bengio, updates the risk landscape annually, and ranges across harms from bias and labor-market disruption to loss of control.[16] Many of the entries on these maps were identified by researchers working inside the frontier companies.

Compared with this crowded risk landscape, the safety frameworks the companies have written for themselves are strikingly sparse. As of mid-2026, the frameworks of Anthropic, OpenAI, Google DeepMind and Meta each track only a handful of categories of risk. And these tend to be capability-driven risks — risks defined by what a model can be shown to do on a test. These include the "uplift" a model could give someone seeking to build biological or chemical weapons, cyber offense, and varieties of AI self-improvement or loss of control (with a single exception, which I will return to below, where manipulation has recently joined the list).[2,7,12,13] Each framework also applies a "severity floor" to a risk or risk category of the kind OpenAI's 2025 revision to its Preparedness Framework made explicit. Meta says something similar, stating that broader risks are addressed through processes "outside of the scope of this Framework".[13] This phrase represents, in effect, a boundary of accountability that is defined by the company itself. Everything below it — including manipulation and persuasion, misinformation, the erosion of human agency, harms accumulating gradually across millions of small interactions, and, notably, nearly every risk these companies pose to themselves through their own cultures, governance and public standing — is below it by choice. These are recognized risks. But they are risks that few if any formal frameworks identify, own, or articulate accountable ways of managing.

This, I would argue, is a serious omission if these "orphaned" risks present a credible threat to an organization, its stakeholders, or society more broadly. However, to be fair to the frameworks' designers, there is a case for focusing on a narrow but deep risk layer rather than trying to be all-inclusive. And it comes down to how finite resources are invested. Concentrating limited safety resources on risks perceived to have the highest levels of severity makes sense; a framework that tried to manage sixteen hundred risks would end up managing none of them. And the frameworks' architects have, sensibly, never claimed completeness. It's also important to recognize that safety frameworks are not the only risk-management tool that these companies have at their disposal: usage policies, trust-and-safety teams, societal-impact programs, and more, all help address parts of the excluded risk territory. But these are discretionary. As with OpenAI's reassignment of persuasion, they carry no public thresholds, accountability, or pre-committed responses. And they can be reorganized or defunded with speed, and without anyone outside the company knowing. In contrast, what the frameworks track is what the companies have committed, in public, to be accountable for. Everything else depends on trust and goodwill — and the record examined below (and my own experiences working with entrepreneurs[17]) shows that under competitive pressure, such commitments tend to become weakened.

What emerges is a risk landscape where the gaps are gaps in accountability. Many of the excluded risks are known and named — in many cases in documents written by employees of the same companies developing risk frameworks. They are just unowned, where ownership means public, pre-committed and versioned accountability rather than a team somewhere having them on their to-do list. Some years ago, working with entrepreneurs facing a similar landscape of recognized-but-unmanageable threats, I started referring explicitly to risks like these as *orphan risks*: risks for which no agreed-on tools, standards or mitigations exist, which no one is accountable for in practice, and which, for that very reason, have a habit of being overlooked and sidelined, even though they may blindside an enterprise later on, or lead to serious societal harm.[18,19] The term was coined specifically in the context of startups and other organizations grappling with emerging and often hard-to-pin-down risks. But it describes the risk landscape faced by frontier AI systems and their developers and users just as well.

Once gaps in the frontier AI risk landscape are seen in terms of orphan risks, an interesting and potentially useful question arises: how do AI risks become orphaned? If it is accepted that many of these risks have been identified or are not hard to identify (while acknowledging that there will inevitably be some emergent risks that have so far eluded identification), rather than just asking what the orphaned risks are, a more revealing question is how they are made. In effect, by what process does a known risk come to be nobody's responsibility?

## 3. How risks become orphans

That institutions (and society writ large) choose which dangers to focus on and which not to — and that the choices reveal as much about how they are organized and operate as they do the objective nature of threats — is generally recognized. In 1982, the anthropologist Mary Douglas and the political scientist Aaron Wildavsky argued that every society selects its risks, ranking some dangers as intolerable and ignoring others — and that you can therefore understand an institution from what it fears. In effect its list of feared dangers reflects its own organization and — in the corporate case — its mission and ambitions, as much as any independent assessment of threat.[20] More than a decade later, the historian of science Theodore Porter showed how institutions under external scrutiny tend to retreat to what can be quantified. Numbers, of course, do not always uniquely capture the essence of what is relevant to decision-making. Professional experience, judgment, and intuition, along with other factors, inevitably come into play as well. But quantitative evaluations — numbers — have a unique power within decision-making ecosystems. Where judgment often has to be taken on trust, numbers can be handed to an outsider, checked, and defended.[21] Such foundations of how risk is assessed, quantified and managed were historically worked out in a landscape comprised of nuclear plants, chemical works and government bureaucracies. And through most of this working-out, much of the process was hidden. But I would argue that frontier AI models and systems, while being an extension of this risk landscape, offer something different: a public, timestamped, versioned record of risk selection in progress. And this is a record not merely of what firms disclose, but of what they commit to manage. Reflecting this, between 2023 and 2026, each of the four developers examined here revised its framework at least once, and every revision is preserved and comparable with its predecessor.

Read in sequence (Table 1), this record can be viewed through the lens of four filters. These can be articulated as four questions that apply to every candidate risk: Can we measure it? Is it big enough? Can we evidence it? And can we afford to keep it? The record suggests that a risk has tended to survive in a safety framework where the answer to all four is yes.

**Table 1. Charting how frontier AI risk commitments changed, 2023–2026**

| Document (trajectory) | Key changes in the public record | Direction |
|---|---|---|
| OpenAI Preparedness Framework, Beta (Dec 2023) → v2 (Apr 2025) | Persuasion removed as a tracked category, reassigned to usage policies and misuse investigations; severe-harm floor made explicit (“death or grave injury of thousands of people or hundreds of billions of dollars of economic damage”); sub-threshold “research categories” added; tracked set focused on biological/chemical, cyber, AI self-improvement.[1,2] | Committed scope narrowed; monitoring tier added |
| Anthropic Responsible Scaling Policy, v1.0 (Sep 2023) → v3.3 (May 2026) | v1.0’s bolded commitment “to pause the scaling and/or delay the deployment of new models” whenever scaling outstripped safety procedures became, through the Feb 2026 rewrite, a discretionary and competitor-conditioned judgment, compensated by periodic Risk Reports and a nonbinding safety roadmap. The v2.0 revision (Oct 2024) had considered persuasion explicitly and set it aside as “not yet sufficiently understood to include in our current commitments”.[7,22] | Commitment weakened; process added; persuasion considered and declined |
| Meta Frontier AI Framework (Feb 2025) → Advanced AI Scaling Framework v2.0 (Apr 2026) | Response to the most severe risk threshold changed from “Stop development” to “Develop with Mitigations’; primary threshold standard loosened from capabilities that would “uniquely enable” a catastrophic outcome to those that would “substantially contribute to” one; loss of control added as a domain; whistleblower protections added; broader risks “outside of the scope of this Framework” in both versions.[13] | Scope widened; response commitments softened |
| Google DeepMind Frontier Safety Framework, v1.0 (May 2024) → v3.1 (Apr 2026) | Harmful manipulation added as a risk domain (v3.0, Sep 2025); sub-critical tracked capability levels added (v3.1). An earlier revision (v2.0) removed the standalone autonomy domain, relocating it to a misalignment section.[12] | Scope widened (the counterexample: voluntary tracking of manipulation is feasible) |
| OpenAI Frontier Governance Framework (May 2026); Anthropic Frontier Compliance Framework (Dec 2025, revised 2026) | Compliance frameworks under California SB 53 (both companies) and the EU Code of Practice (OpenAI). OpenAI’s covers harmful manipulation — a category its safety framework does not track — describing its approach as exploratory; Anthropic’s initially mirrored its safety-framework categories, then added nascent harmful-manipulation tiers in an early-2026 revision, ahead of European enforcement.[3,8,9] | Scope widened where regulators set the list (the “safety differential”) |

**Can we measure it?** There are strong indications that a risk earns a place in a frontier framework only if it can be operationalized, or turned into an evaluation with a threshold; in effect a capability threshold. This is a pragmatic decision based on established approaches to risk management. A capability threshold relies on what a model can be shown to do on a test, as opposed to a risk threshold, which considers how likely a harm is. And the first is far easier to evaluate reliably.[23] As well as representing reasonable engineering judgment, such an approach is also a concession to the reality that what the frameworks track follows what their instruments can measure. What happened to persuasion in the case of OpenAI illustrates this filter at work — though not, as we will see, working in isolation. OpenAI’s stated reason for removing this category was one of fit: persuasion-type risks, the revision explained, did not meet its criteria for tracked categories — criteria that required a tracked risk to be both severe in its potential harms and measurable in practice.[2] Yet the risk itself was demonstrably not beyond measurement. Within months of the removal, a large experimental study in the journal Science demonstrated that conversational AI systems measurably shift political beliefs — a persuasion-based effect.[24] What persuasion lacked in this case was not measurability, but measurability in the accepted idiom — a pre-deployment capability evaluation with a clear threshold that could be defended in a court of law.

**Is it big enough?** Severity floors make sense as triage when assessing and managing risk. But they also carry hidden consequences. A risk that emerges gradually, distributed across millions of small interactions, will rarely trigger them. The AI philosopher Atoosa Kasirzadeh draws a useful distinction here between "decisive" pathways to AI catastrophe — a single dramatic event — and "accumulative" ones in which harms compound below the threshold of any single incident, until a societal point of no return is crossed.[25] A framework built of capability evaluations that are run against a catastrophe floor will tend to be, by its own construction, insensitive to the accumulative pathway. Few, if any, of the evaluations the AI safety frameworks considered here are designed to trigger action based, for instance, on the slow erosion of trust, of human agency, or of the technology's social license — the informal public permission on which its continued operation depends. And this, in turn, opens the door to the "big enough" filter excluding potentially relevant risks.

**Can we evidence it?** Under scrutiny, risk assessment and management drift toward what can be shown and measured. The sociologist Michael Power called the result of such drift the "audit society": a society comprising organizations that answer demands for control with rituals of verification, so that — when applied to risk — the deliverable becomes the documentation rather than effective risk management.[26] This is a pattern that is strongly indicated in the frontier frameworks considered here. When independent researchers, for instance, scored the published frameworks against sixty-five criteria covering how well they identify, analyze, treat and govern risk, the strongest framework earned only around a third of the available points, while the median company earned fewer than one in five.[27] Reading the frameworks, it is not hard to see why: they are at their strongest where demonstration of action is easy through published thresholds and named evaluations, and at their weakest where it is not. A framework, it turns out, can be an excellent exhibit, and a weak instrument, both at the same time.

**Can we afford to keep it?** The fourth filter here is different to the previous three as it cannot be identified directly in any single document. Rather, it shows itself only over time through what happens to articulated commitments as they come under pressure. And the pattern seen in the records is that, when a commitment starts to look as though it might actually constrain a company, it tends to soften. Anthropic's original scaling policy of 2023 for instance contained a bolded, unconditional commitment "to pause the scaling and/or delay the deployment of new models" whenever scaling outstripped safety procedures.[7] Compare this to the comprehensive rewrite of February 2026, which turned that unconditional pause into a discretionary one which is conditioned on what competitors do, and not safety in isolation.[7,22] Anthropic's Holden Karnofsky, who led the rewrite and was careful to note he was writing in a personal capacity, defended the change on the grounds that it is no good getting responsible actors to slow down unilaterally while others press ahead — underlining the influence of corporate success (including within a global market) on risk strategies.[22] Meta's revision the same year was franker still: it changed the required response to the company's most severe risk threshold from "Stop development" to "Develop with Mitigations", and loosened the threshold's primary standard from capabilities that would "uniquely enable" a catastrophic outcome to those that would "substantially contribute to" one.[13] Each of these changes, I would suggest, was locally reasonable, publicly logged and individually defensible, and yet resulted in diminishing or orphaning categories of risk — potentially to the detriment of enterprises, key stakeholders, users, and society more broadly. And that is what makes the pattern worth taking seriously.

This is a pattern that is not unique to frontier AI models and systems, and is something that is highlighted in the roots of the 1986 Challenger disaster as a salutary example. The sociologist Diane Vaughan spent years reconstructing the disaster from NASA's paper trail, and what she found was a sequence of individually justified acceptances of deviations from standard practice, each one resetting the baseline against which the next was judged, until the organization had normalized what would once have been intolerable.[28] Vaughan was writing about hardware in this case — eroded O-rings on a solid rocket booster — but the mechanism she described operates just as readily on written commitments. It is also what Jeffrey Abbott and I refer to as *values drift* in our book *AI and the Art of Being Human*: not dramatic betrayals but "the small yes that makes the next yes easier", until an organization is agreeing to things that would have alarmed it a year before.[29]

The exploration here of how risks become orphans is, not surprisingly, something of an oversimplification. I would argue that it is informative nevertheless. However, it is worth addressing two potential points of concern. The first is an assumption that the direction of travel around risk that leads to risks being orphaned is universal, when it is clearly not. Google DeepMind, the exception flagged earlier, moved the other way, adding an avowedly exploratory harmful-manipulation domain to its own framework in 2025.[12] However, what this example establishes is helpful in that it demonstrates feasibility when it comes to addressing orphan risks. Based on DeepMind's actions, tracking an orphaned risk voluntarily is something a frontier framework can evidently do. And this implies that exclusion elsewhere is — at least in some cases — a choice rather than a necessity. The second potential point of concern is that the assessment above implies cynical motives. However I would argue that it does not. Many of the people writing frontier AI safety frameworks have spent their careers trying to make powerful technologies safer, and I have no reason to doubt that the frameworks address what their authors believe to be relevant and important. But sincerity almost always operates inside an incentive field — and in particular the competitive environment every one of these companies inhabits. And that field has a tendency to reward each small, locally defensible softening, regardless of what anyone intends. Here I would suggest that sincere people, under the constraints of productivity and competition, reasoning one reasonable compromise at a time, produce the same sort of drift that intentional bad actors might create on purpose. And this is itself a form of potentially orphaned risk — one that is identifiable, but not formally recognized or acted on. This is why the patterns observed here do not necessarily need "bad actors" to explain them. It is also why remedies aimed at sincerity, such as exhortation or public shaming, are unlikely to have the desired impact. In effect, if competition is the driving force, remedies have to change what competition rewards — including consensus norms, rules, and costs that land on every organization at once, rather than appealing to any one organization's implicit values.

This is where it's worth coming back to the account I began with of OpenAI and persuasion as a risk. If we set a company's safety framework beside its compliance framework, the gap between them — call it the *safety differential* — points to something specific: the difference between the risk aperture a firm selects for itself, and the one which is selected for it. That OpenAI's omission of persuasion as a risk domain was a choice, not an impossibility, is demonstrated by its own compliance framework covering closely overlapping territory the moment regulators required it.[2,3] Here, the safety differential is an imperfect instrument in that the compliance side tracks whatever regulators list, and it exists only where a regime applies. But it does help highlight the dynamic between the risks that an organization elevates, and those that it's incentivized to elevate (especially through enforcement). Through this lens, a risk category that a company must answer for anyway becomes cheap to own voluntarily, and so such categories should begin migrating into the safety frameworks as enforcement arrives (thus reducing the differential). The sharpest test here comes from Europe, where the timing for compliance has already been fixed: obligations required by the EU have applied since August 2025, but fines for breaching them only begin in August 2026. California's statute, already in force, sets no comparable deadline. Here, Anthropic's early-2026 addition of manipulation tiers, made before any enforcement, may be a first sign of that migration. If, instead, the safety differential is seen to persist — say, past 2028, which allows for a revision cycle or two after enforcement begins — this would indicate the safety frameworks are insulated from the compliance function altogether, and that the differential may not to work as a migration mechanism. That would be a more troubling finding, and one that would count against the incentive-driven account argued here — though not against the case that risks are being orphaned.

Could regulation, then, simply close the gap identified here? I must confess that I am not optimistic, and the reasons can be found in the documents themselves. Compliance coverage is jurisdiction-bound and politically contingent, and the new instruments that are emerging largely inherit the frameworks' own aperture: California's statute for instance confines its mandated disclosures to catastrophic risk, narrowly defined. And OpenAI's compliance treatment of manipulation is, by the company's own description, exploratory — with a far thinner risk assessment and management machinery than the safety frameworks apply to their chosen risks.[3,4] Recall, too, that the statutes were built partly on the frameworks' own patterns, and so the aperture has a tendency to travel with them. And, I would argue, current statutes barely reach the risks that have arguably cost these companies most — the ones living inside their own missions, cultures and relationships of trust. For those, I suspect that the fix cannot come from statute alone. Rather, it has to come from rethinking how the companies themselves define and approach risk. And this leads to how the framework of risk innovation might be applied.

## 4. Risk as a threat to value

A useful place to start when considering how risk is defined and approached is the four filters introduced earlier. Three of these can be traced back to a definition of risk that the frameworks share: risk as the probability of a specified, severe harm event. Under this definition, if a harm cannot be specified in advance, there is nothing to build an evaluation around, and any associated risk cannot become part of the framework. This is the measurement filter. On the other hand, if the harm arrives as a myriad small losses rather than one severe event, it will struggle to cross the severity floor — the severity filter. The next filter is the evidence filter, and this is more complex because the push toward what is demonstrable comes from outside scrutiny rather than from the definition itself. Yet the definition of risk used still guides what counts as a demonstration of risk. And here frameworks lean strongly toward capability-based evaluations as a proxy for risk, rather than calibrating against risk directly. As such, a framework built on risk as the probability of a specified severe harm event is not necessarily being distorted when it excludes what is considered to be unmeasurable, gradual and hard to find evidence for. Rather, it is actually working as designed. It's just that the design itself may be flawed.

Compared to the previous three filters, the fourth filter's relationship with risk is slightly different. Competitive cost has relatively little to do with how risk in a conventional sense is defined. Rather, it relates to whatever commitments exist, however they were conceived. Yet once an organization's own competitive standing is recognized as representing value that's at stake, competitive cost stops being an external force and becomes one more threat to value. And this includes potential threats to mission, users' trust, and license to operate. The fourth filter, then, is not necessarily removed by the value lens, but is absorbed into it.

Given this, is there a case to be made that framing risk in the context of frontier AI as probability-of-harm too narrow a foundation for managing potential harms? Quite possibly. Of course, the idea that probability-of-harm is too narrow a foundation for risk management is not new. And some of the existing alternatives get close to what could be useful here. Mainstream enterprise risk management, for instance, moved beyond pure probability-of-harm definitions a number of years ago. In this context, ISO 31000, the international risk-management standard, defines risk as the "effect of uncertainty on objectives", a definition inherited verbatim by the AI risk-management standard that descends from it.[30] On the surface this is close to a definition of risk that I propose below, and is certainly a useful step toward a more productive definition of risk. The difficulty lies, though, in whose objectives count as this definition is applied, and in what gets admitted as an objective. In practice, the objectives within enterprise risk management approaches are the enterprise's own, and usually conventionally accounted for — revenue, operations, reputation as the market prices it, and so on. In this way, enterprise risk management approaches are, as a result, capable of naming risks that the safety frameworks tend to orphan (the securities filings quoted earlier indicate as much). But they attach little to those risks beyond disclosure, including no committed management, and no standing for any value beyond the organization's own.

Coming from the other direction, a half-century of scholarship on the social nature of risk has insisted that publics and their values belong inside risk appraisal, not outside of it. And one strand of this tradition matters in particular here. The risk analyst Roger Kasperson and his colleagues showed how harms amplify through social response; how a seemingly minor technical event can ripple outward into losses of trust, legitimacy and market standing that dwarf the original damage.[31] Here, work on responsible research and innovation for well over a decade has distilled these and associated insights into frameworks and practices for technology developers. And while this is now a broad and diverse field, the early work of Stilgoe, Owen and Macnaghten can be summarized as: anticipate consequences before they arrive; include the people who will bear them; reflect critically on your own assumptions, values and framings; and respond — actually change course — when there are indications that something is wrong.[32]

Here I would also be remiss if I did not acknowledge that the EU General Purpose Code of Practice supports innovation in AI safety and encourages "providers of general-purpose AI models with systemic risk to advance the state of the art in AI safety and security and related processes and measures".[6] And yet, despite governance steers, scholarship, and practice-based frameworks around alternative approaches to risk, frontier AI risk frameworks still veer toward relying on conventional risk definitions. This, I suspect, is partly because the societal tradition speaks a language that is ill-suited to how fast-moving firms make decisions. Some years ago, Elizabeth Garbee and I explored why responsible innovation frameworks struggle inside entrepreneurial cultures, drawing on my experience working with entrepreneurial engineers.[17] What we found was not indifference, but something else. Innovation cultures that sincerely want to do good nonetheless tend to reject frameworks that arrive as top-down obligation — and respond, often enthusiastically, to framings built around the creation and protection of worth. The lesson that has stayed with me ever since is that if you want a fast-moving organization to attend to a risk, you do not hand it a compliance duty; you show it a threat to something it values. Frontier AI labs — mission-driven, often allergic to imposed process, and rarely short of conviction in their own exceptionalism — align closely with the culture we described.

This is the gap that work around *risk innovation* was built to fill. Its seeds were planted in 2013, while I was teaching entrepreneurship students at the University of Michigan who faced a bewildering landscape of hard-to-quantify social and political risks that none of their business tools addressed. It took institutional form when I launched the Risk Innovation Lab at Arizona State University in 2015, which was where formative concepts geared toward navigating novel risks from emerging technologies began to come together.[33] And it matured between 2017 and 2020 as the Risk Innovation Accelerator, later the Risk

Innovation Nexus: initiatives that built and piloted a toolkit and risk navigation resources with time- and resource-constrained entrepreneurs in mind.[19] The design principles we developed and leveraged were driven by utility — tools had to be simple and intuitive, they could not afford to demand any heavy time investment, and they needed to complement a company's existing risk management approaches — because the people they were built for had little time and even less money to invest in risks that did not fall into conventional categories, but were still a threat to what they were trying to achieve.[19] The framework has since been applied a number of times, most fully in a multi-organization study mapping how sixteen partner organizations in a biopreservation research ecosystem perceived value and orphan risks,[18] and it shapes much of my own approach to AI risk.

At the core of the risk innovation approach is an operational definition of risk that creates a pragmatic route to adopting and addressing orphaned risks: treating risk as a *threat to value*. This does not abandon the idea of risk as involving the probability of harm. Rather, it widens what counts as harm — from a specified catastrophic event to an impact on anything that carries worth or value. Value, in this framing, can be tangible, such as health, security or revenue; or intangible, such as trust, autonomy or dignity; or even aspirational — the positive future an organization exists or strives to bring about. Importantly, value (or worth) within the context of the risk innovation framework is not just held by the enterprise, but by its key stakeholders: its investors, customers and the communities it impacts.

To illustrate the shift that this seemingly simple reframing of risk brings about, consider managing the probability of a specified harm to dignity. In this case, the conventional machinery of risk has little or nothing to run on. But consider instead who (or what) holds dignity in a given situation, what threatens this, and what protecting it would look like; and a threat to dignity seen as a threat to value or worth becomes something that can be acted on — even though nothing has been quantified. And here the definition is a deliberate fusion of approaches and framings: it keeps the strategic, value-protecting orientation that makes enterprise risk management adoptable, while also widening the circle of those whose value counts. And its practical value comes from a coupling that might be summed up as *your risk is my risk*: the idea that threats to what your *stakeholders* value convert, through the amplification dynamics that Kasperson describes, into threats to what *you* value — through channels such as public backlash, the flight of talent, litigation, regulation triggered by lost trust, and more.[17,31]

Within this frame, orphan risks are not simply one more entry in the already long list of AI risk taxonomies. They are a way of identifying and naming potentially neglected risks that are nevertheless important: threats that are recognized but unowned because no established tools or frameworks exist to address them effectively. The operational version of this — developed under the umbrella of the *Risk Innovation Nexus* — groups eighteen such risks, among them loss of agency, damage to organizational values and culture, and erosion of public trust, into three domains: social and ethical factors, unintended consequences of emerging technologies, and organizations and systems.[19] It pairs this map, or risk landscape, with a deliberately lightweight set of operational tools, including the Risk Innovation Planner, which asks users to identify a few areas of value for each stakeholder group, consider which orphan risks threaten them, commit to a handful of small actions that are completable within a few weeks, and then repeat.[19] While this is just one implementation of the risk innovation framework, it's worth mentioning as the simplicity and ease of using the Planner are indicative of how intentional design decisions have been used to connect theory to practice within the risk innovation framework. Here, it's well known that practices tend to survive inside fast-moving organizations when they return visible value quickly and augment what already exists. As a result, the Planner and other risk innovation implementations are intentionally designed to support high-value and low-cost practices for people and organizations with little time and limited resources. And here, they potentially offer frontier AI a framework and a set of tools that enables the "de-orphaning" of critical risks.

## 5. What the value lens reveals

Given this, what happens when the way risk is defined and framed changes? Here it's worth returning to the four filters, approaching them through the lens of risk-as-threat-to-value. Through this lens, the first three filters rapidly lose their tendency to exclude certain risks, as value can be named, mapped and watched — even where it cannot be measured. And as a consequence, threats such as the slow erosion of trust or agency can begin to be treated as a risk in its own right. The fourth filter — competitive cost — is, as noted earlier, absorbed by the risk-as-threat-to-value lens. And that changes how an organization approaches its commitments. For instance, a company weighing whether to keep a commitment might consider what it would lose by breaking it. Framed as what is, in effect, a tax on competitiveness, a commitment protects nothing the company can point to, and it will always be vulnerable to being modified or removed under pressure as a result. Yet when framed as protecting something of value or worth that the company demonstrably depends on, intangible as this might be — its mission for instance, or its talent, or license to operate — the same commitment is quickly reframed as something the company knows it needs, and that must be defended against threats.

Of course, a more practical test here of such a reframing of risk is to ask whether a redefinition like this would have helped avoid real damage which existing frameworks previously missed. Here, it's worth considering some of the events that have arguably damaged frontier AI companies most since 2022. When Meta demonstrated Galactica for instance (a large language model for science) in late 2022, the public demo lasted three days. What sounded its death knell was a conflict between the system's fluently confident errors and something the scientific community values deeply: credibility.[34] Through a value lens, the risk would have been visible and hard to orphan: a threat to community trust and to the perception of the enterprise. Yet as it was, this risk was overlooked as it sat squarely in territory the conventional frameworks did not cover.

Another example is seen in the OpenAI board crisis of November 2023, in which the company's nonprofit board dismissed its chief executive Sam Altman, citing a loss of confidence in his candor — only to reinstate him days later after nearly all of the company's employees threatened to leave. This was, at one level, precipitated by threats to organizational values. A governance structure built to protect an aspirational mission collided very publicly with commercial reality, and very nearly destroyed the company it was designed to safeguard in the process.[35] In this case, the risk was not central to the models being developed, but was integral to the ecosystems within which they were being developed. And the safety team's departures that followed in 2024 made the cost of that threat to value concrete. Jan Leike, who had co-led the company's work on aligning future systems with human intent, captured the problem in a single sentence as he left, writing publicly that "safety culture and processes have taken a backseat to shiny products".[36]

A third example here is litigation that is beginning to put user wellbeing — a value that few frontier AI frameworks track — onto the legal record. This is perhaps seen most prominently in a wrongful-death suit brought against OpenAI by the family of a California teenager who allegedly took his own life under the influence of ChatGPT.[37] But this is just one instance of a growing movement toward communities using legal action to push back against the impact of AI and associated technologies on wellbeing. And while such actions do not fit neatly into AI safety frameworks, they nevertheless represent a threat to value that could have substantial consequences for frontier model development and use.

These examples are, of course, anecdotal, and serve more to illustrate the utility of approaching risk as threat to value with frontier AI than as evidence of its necessity. And the companies concerned in each case managed to absorb each threat (although it's still too early to gauge the long-term consequences in the case of user wellbeing) and, by market measures, continued to thrive. But the concern here is that indications of thriving are an artifact of assessing risk within a relatively short time window, and with a threshold of catastrophe rather than incremental harm. In contrast, a risk-as-threat-to-value lens would suggest that potential damage accumulates over time, is easy to overlook in the short term, and emerges from risks that are not codified within existing frameworks — no owners, no indicators, no registers, and nothing whose removal would even be noticed.

Beyond these examples, there is another aspect of frontier AI safety that the value lens reveals that I think is worth paying attention to, and that further supports the adoption of orphan risks. The founding documents of the companies considered here — documents that precede the safety frameworks — are accounts of aspirational value. OpenAI's charter, for instance, promises to ensure that artificial general intelligence "benefits all of humanity".[38] It is easy to dismiss such commitments as branding. But they are perhaps better understood as assets — the basis of talent attraction, public trust, and regulatory goodwill for instance — and, like any assets, they can be spent. Approached this way, OpenAI's framework revisions of 2025 and 2026 are a public, self-published record of those assets being drawn down under competitive pressure, one defensible "softening" at a time. Here, I would argue that a company that is genuinely tracking threats to its own aspirational value would treat its framework changelog as a leading indicator that the company is drifting, in public and by increments, from the mission it was founded on.

Looking forward, the value lens is useful as a pointer to where the next blindsides may occur, and in particular the places where deployment is racing ahead of anyone owning potential risks. Three areas in particular stand out here as being worthy of attention through the lens of risk as threat to value. The first is emotional reliance. As AI companions and assistants scale into hundreds of millions of lives, dependence on them is likely to stop being an outlier, and is increasingly likely to become a population-level phenomenon. This is a phenomenon that is already emerging and leading to litigation and legislation. And yet it is not addressed directly by any existing safety framework bar, possibly, DeepMind's new "harmful manipulation" level — and this targets mass manipulation rather than personal emotional reliance.[12] The second is the erosion of epistemic agency: people's control over what they come to believe as frontier AI models and systems become more prevalent. Persuasive, personalized systems now increasingly mediate what people read, consume, and are exposed to through various channels. And I have argued elsewhere that fluent, endlessly obliging AI may function as a kind of cognitive Trojan horse — bypassing the vigilance we instinctively apply to human persuaders because it carries none of the cues that trigger it.[39] And researchers are already documenting a tendency to adopt AI outputs with minimal scrutiny, overriding both intuition and deliberation, through what has been called cognitive surrender.[40] And the third is the developers' own safety culture, which is already a source of internal values-based conflicts, and under growing pressure as competition and political pressure compress timelines and change the operational rules of the game.

These are just three areas where a risk-as-threat-to-value lens can help reveal risks that are easy to ignore, are poorly addressed in current safety frameworks, and yet are nevertheless likely to be consequential — there are no doubt many more.

**Table 2. Orphan risks in frontier AI: documented manifestations and coverage as of mid-2026**

| Orphan risk (domain) | Documented frontier AI manifestation | Safety-framework coverage | Value at stake (holder; kind) |
|---|---|---|---|
| Perception — how the enterprise and its products are seen (social & ethical) | Galactica demo withdrawn within three days amid credibility backlash, 2022.[34] | None | Trust in the enterprise and the science it draws on (enterprise, community; intangible) |
| Organizational values & culture (organizations & systems) | OpenAI board crisis, 2023; senior safety resignations, 2024.[35,36] | None | Mission integrity, talent (enterprise; aspirational) |
| Loss of agency (unintended consequences) | Persuasion dropped from OpenAI's voluntary tracking, 2025, amid published evidence of scaled machine persuasion.[2,23] | Google DeepMind only (exploratory), as of mid-2026.[12] | User autonomy and epistemic agency (customers, community; intangible) |
| Social trends — changes in how people live with the technology (social & ethical) | Companion-chatbot reliance; wrongful-death litigation.[37] | None | Customer wellbeing; social license (customers, community; tangible and intangible) |
| Reputation & trust (organizations & systems) | AI-driven reputational harm a dedicated risk factor in securities filings — disclosed to investors, unowned in frameworks.[10,11] | None | The substrate on which all other value conversion runs (all groups; intangible) |
| Health & environment (unintended consequences) | Energy and water footprint of compute; impacts diffuse and borne by communities.[16] | None | Community environmental value (community; tangible) — harm converts to enterprise cost late, if at all, typically through organized backlash |
| Social justice & equity (social & ethical) | Data-labor conditions in annotation supply chains; harms to non-users.[16] | None | Dignity and fair treatment (community; intangible) — conversion channels exist but run slowly and unevenly |

Yet such a value-based approach does have its limitations, and these potentially fall hardest on the people with the least leverage over what the organizations driving AI development are doing. The your-risk-is-my-risk coupling that is at the core of the risk innovation framework runs through what might be called conversion channels — backlash, litigation, talent, and regulation for instance — and those channels are not equally open to everyone. History indicates that they are rarely closed entirely: communities have made firms feel harm through movements before, from the consumer revolt against genetically modified food, for instance, to today's local resistance to data centers. And citizen pressure has a way of arriving eventually as legislation. But those channels often work slowly, are blunt instruments, and involve a considerable time lag. As a result they tend to convert harm into enterprise cost only after the harm is done — and unevenly at that. Data workers in annotation supply chains for instance, or communities carrying the environmental costs of compute, or people affected by systems they never chose to use: for them, a value lens operated by an enterprise may register the risk only when a movement forces it to (Table 2 flags some such cases). This is certainly the case where conventional risk and safety frameworks dominate organizational decisions. And yet, this is simply another form of orphaning risks that will potentially come back to bite enterprises in the future because they failed to take seriously threats to value to the organization *and* to its stakeholders and the communities it impacts. And because of this, the risk innovation framework — and approaching risk as threat to value within an interconnected system of actors — opens the way for frontier AI developers to actively leverage value within such systems to make decisions that avoid future harms that conventional approaches may overlook.

## 6. What would change in practice

Based on what is known of the emerging AI risk and safety landscape, what leading companies' internal documents show, and what regulator-aligned documents reveal, this assessment suggests that orphaned frontier AI risks could potentially create vulnerabilities for developers, users, and society writ large — and, by inference, for economic growth and national security — and that reframing risk as a threat to value could help mitigate or "de-orphan" some of these risks. The toolkit that has already emerged around the framework of risk innovation provides AI developers and others with something that current frontier frameworks overlook: a structured and responsive way of thinking about risk — together with a set of tools — that were built to navigate easy-to-overlook yet critical risks in ways that are adaptable and scalable to different users' needs.[19] A frontier AI enterprise, for instance, could readily adopt the quarterly practice outlined in the Risk Innovation Planner as it stands, or adapt it for their specific circumstances. The tools associated with the risk innovation approach are freely available for using and modifying. And the full set of tools and resources were designed to complement existing risk machinery rather than replace it. And as the worked example shown in Box 1 illustrates, implementation of the framework is likely to be relatively low cost, with potentially high returns for developers, adopters, policymakers, users, and beyond.

That said, what the toolkit does not supply by itself is public accountability (although this is integral to the landscape defined by the risk innovation approach). And frontier AI needs this, as it is being developed and deployed by companies whose private risk selections have become, as I argued at the outset, a de facto layer of public governance. Here, two disclosure documents would help extend the use of the risk innovation framework into such a role. The first is an *orphan-risk register*: a standing, public annex to the risk reports some developers have already committed to publish.[7] Each entry would record a risk the company considered and decided not to manage, and give the reason — it could not be quantified for instance, or it fell below the organization's severity floor, or managing it was judged too costly under competition. The second is an *aperture log*: a short statement accompanying each framework revision that records what was scoped out and why. Neither document would require the company to manage anything new — although one would anticipate that, over time, orphaned risks would, in effect, be adopted. What each would do, though, is put the company's scoping decisions where others can monitor and respond to them, thus introducing an additional layer of accountability. Here, a de-listed risk that has to be explained in public is a de-listed risk that regulators, researchers and employees can ask about and hold an organization to account over. This then would become a lever on the fourth filter described above — the one asking "can we afford to keep it?" — that no redefinition of risk alone could supply. In effect it places a cost on decisions or walk-backs that must be explained.

Whichever mapping practice lies behind such an orphan-risk register, the framing of risk innovation would require that its community-facing entries come — at least in part — from structured engagement with people outside the firm, as this is where true stakeholder value emerges. And here it would make sense for the register to note any objections received over time, and how the map changed as a consequence.[18] And this is important within the framing of a risk landscape comprised of orphan risks, as a register whose map never changes, however hard outsiders push on it, would suggest the engagement is theater — and thus another (in this case self-generated) risk to be navigated.

Of course, it could be argued that such a register is just the kind of artifact the audit society as described earlier would co-opt — produced through ritual and process, reassuring by its very candor, but changing nothing.[26] Yet if implemented well, there is no reason why it could not rise above such a pathway, low-resistance as it may be. A register's value would hang not on its existence, but on the record of revisions it captures. And this is measurable, or at least observable. If orphan-risk registers are adopted, and then captured as a form of safety theater, the capture will presumably show up in the record, and thus allow the adopter to be held to account.

Within this framework there is also a role for regulators. Rather than mandating coverage of every risk — a requirement that would be unworkable in practice — they could require companies to disclose how they select the risks they cover. It's a move that would, at the very least, help identify what is being orphaned, and would more likely encourage greater reflexivity around what is adopted. California's transparency reports and the EU code's documentation requirements are existing vehicles into which such requirements — associated with an orphan-risk register and an aperture log — could be folded at little additional cost. And such a move would extend actions in a direction that both regimes have already taken.[4,6] Importantly, regulation of this kind would not necessarily need to decide which risks matter. It would simply ensure greater visibility around who is deciding what matters, and on what grounds.

## 7. Taking stock

The heart of this paper's argument — that frontier AI's risk apparatus selects for the quantifiable, the catastrophic, the auditable and the competitively affordable, that the selection of relevant risks tightened between 2023 and 2026, and that such an approach introduces risk and safety vulnerabilities for developers, users, and society more broadly — rests on decades of scholarship on how institutions choose their risks, together with the public records summarized in Table 1. The risk innovation framework — including reframing risk as a threat to value and categorizing important but easily sidelined risks as orphan risks — is presented as one way of addressing vulnerabilities here — not as an alternative, but as an augmentation of current risk and safety frameworks, governance approaches, and management strategies.

That said, what is presented here is an analysis of an emerging risk landscape and a potential response that I would argue is defensible, but has yet to be shown to be useful in practice. And here, it's worth considering three tests that can help reveal the degree to which this assessment might apply in specific situations: (1) whether risks de-listed from safety frameworks generate (or potentially generate) incidents and costs at rates comparable to tracked ones; (2) whether the differential between safety and compliance frameworks narrows (or is likely to narrow) from the voluntary side once European enforcement begins in August 2026; and (3) whether adopting an orphan-risk register changes (or has the potential to change) what subsequent framework revisions cover.

To be clear, nothing here argues that the catastrophic-capability apparatuses that are already in place should be loosened. Rather, the paper argues that a single safety layer is currently being asked to effectively stand in for two, and that the second layer — the one that would allow threats to value to be navigated effectively — is missing, or at least diminished. And here there is an urgency to both ensure that this layer is present and robust, and to provide opportunities for enterprises to succeed through transforming vulnerabilities associated with orphaned risks into advantages that come from being able to navigate a complex risk landscape with open eyes.

This analysis started with the safety differential — what frontier AI developers recognize as potential risks as opposed to what they are publicly accountable for — and argued that what becomes sidelined is not necessarily what is low risk, but what does not fit within existing risk and safety approaches. It then introduced the risk innovation framework as a way of making such sidelined risks visible, and making a treacherous risk landscape more readily navigable. Whether such a reframing of risk and safety is necessary, or advisable is, of course, still open to further testing and exploration. But as a final observation, it is worth noting that, on the record of the past four years, the risks most likely to blindside frontier AI are not the ones its institutions are currently watching. Rather, they are the ones its institutions have organized themselves not to see. And running blind has never been a particularly good risk management strategy — especially where the stakes are high, as is increasingly the case with emerging frontier AI models.

## AI use statement

The research question, the argument architecture, key concepts — including risk innovation framework and orphan risks, drafting, final editing, and all editorial judgments in this paper, are the author's. Large language model tools (Anthropic's Claude Fable 5, used within Claude Code in ultracode mode) were used, under the author's close direction, to research the documentary record, to verify claims and citations against primary sources, and to develop preliminary drafts, which were subject to multiple iterations of author critique and annotation. The author takes full responsibility for all content, claims and citations.

### Box 1. Terms, and a worked example

**Risk as a threat to value:** Replaces "probability of specified harm" as the working definition of risk within the risk innovation framework.

**Value:** Anything that bears worth — tangible, intangible or aspirational (value that does not yet exist, such as that captured in an organization's mission) — held by one of four stakeholder groups: the enterprise, its investors, its customers and its communities.

**Orphan risks:** Recognized (or recognizable) threats to value that remain sidelined and essentially un-owned within conventional risk frameworks. No agreed tools to address them; no implemented standards or mitigations; and no one who is publicly accountable for them.[18,19]

**Risk innovation:** The practice of conceptualizing, approaching and navigating risk in innovative and, as a result, productive ways — particularly in the context of emerging technologies — and of leveraging this to navigate complex risk landscapes.[33]

**A worked orphan-risk register entry:** (This is illustrative only, and indicative of how it might be used in a developer's periodic risk report. The format is adapted from the Risk Innovation Planner[19]):

*Risk*: Emotional reliance on companion-mode assistants (orphan risk categories: social trends; loss of agency).

*Value threatened*: Customer wellbeing and autonomy (customers); youth mental health (community); litigation exposure and trust (enterprise).

*Current status*: The subject of company research, and increasingly of litigation, yet not tracked in any frontier safety framework; company attention is discretionary and not tracked.

*Why not managed within the framework*: No agreed metrics; harms accrue per user, and thus lie below the established severity floor.

*Register action*: Consolidate existing internal signals into one named, board-visible indicator.

*Owner*: [Named executive].

*This quarter*: Commission an external review of the emerging clinical evidence; add reliance-pattern flags to usage dashboards; brief the board risk committee.

*Review*: Next quarter.

The hypothetical entry above does not commit the organization to quantify, or even to mitigate the risk. Rather, it commits it to consider the risk, on the record, and thus converts ad hoc attention into a disclosed and owned line-item whose removal would itself be visible and send a signal.

## References

1. OpenAI. *Preparedness Framework (Beta)* (OpenAI, 18 December 2023); https://cdn.openai.com/openai-preparedness-framework-beta.pdf (accessed 3 July 2026).
2. OpenAI. *Preparedness Framework, Version 2* (OpenAI, 15 April 2025); https://cdn.openai.com/pdf/18a02b5d-6b67-4cec-ab64-68cdfbddebcd/preparedness-framework-v2.pdf (accessed 3 July 2026).
3. OpenAI. *Frontier Governance Framework* (OpenAI, May 2026); https://cdn.openai.com/pdf/e37d949b-8c9f-4d76-b99e-4272f4631a7e/openai-frontier-governance-framework.pdf (accessed 3 July 2026).
4. California State Legislature. Senate Bill No. 53, Transparency in Frontier Artificial Intelligence Act, Stats. 2025, ch. 138 (2025); https://leginfo.legislature.ca.gov/faces/billTextClient.xhtml?bill_id=202520260SB53 (accessed 3 July 2026).
5. Regulation (EU) 2024/1689 of the European Parliament and of the Council of 13 June 2024 laying down harmonised rules on artificial intelligence (Artificial Intelligence Act). *Off. J. Eur. Union* L 2024/1689 (12 July 2024); https://eur-lex.europa.eu/eli/reg/2024/1689/oj (accessed 3 July 2026).
6. European Commission. *General-Purpose AI Code of Practice* (European Commission, 10 July 2025); https://digital-strategy.ec.europa.eu/en/policies/contents-code-gpai (accessed 3 July 2026).
7. Anthropic. *Responsible Scaling Policy*, versions 1.0 (19 September 2023) to 3.3 (26 May 2026); changelog and archived versions at https://www.anthropic.com/rsp-updates (accessed 3 July 2026).
8. Anthropic. Frontier Compliance Framework. *Anthropic* https://trust.anthropic.com/resources?s=zgt6rgfecy1eo7vicr064b&name=anthropic-frontier-compliance-framework (June 2026) (accessed 4 July 2026).
9. Schaffelder, M. & Murray, M. Analyzing SSF requirements in the GPAI Code of Practice: a case study using Anthropic's Frontier Compliance Framework. *SaferAI* https://www.safer-ai.org/analyzing-ssf-requirements-in-the-gpai-code-of-practice-a-case-study-using-anthropics-frontier-compliance-framework (20 March 2026) (accessed 3 July 2026).
10. Meta Platforms, Inc. *Annual Report on Form 10-K for the Fiscal Year Ended December 31, 2025* (US Securities and Exchange Commission, filed 29 January 2026); https://www.sec.gov/Archives/edgar/data/1326801/000162828026003942/meta-20251231.htm (accessed 3 July 2026).
11. Alphabet Inc. *Annual Report on Form 10-K for the Fiscal Year Ended December 31, 2025* (US Securities and Exchange Commission, filed 5 February 2026); https://www.sec.gov/Archives/edgar/data/1652044/000165204426000018/goog-20251231.htm (accessed 3 July 2026).
12. Google DeepMind. *Frontier Safety Framework*, versions 1.0 (May 2024) to 3.1 (April 2026); v3.0 announcement and version archive at https://deepmind.google/blog/strengthening-our-frontier-safety-framework/ (accessed 3 July 2026).

13. Meta. *Frontier AI Framework, Version 1.1* (Meta, 2025); https://web.archive.org/web/20250311152132/https://ai.meta.com/static-resource/meta-frontier-ai-framework/ (archive of 11 March 2025); and *Advanced AI Scaling Framework, Version 2* (Meta, 2026); https://ai.meta.com/static-resource/Meta_Advanced-AI-Scaling-Framework-v2 (accessed 3 July 2026).
14. UK Department for Science, Innovation and Technology. Frontier AI Safety Commitments, AI Seoul Summit 2024. *GOV.UK* https://www.gov.uk/government/publications/frontier-ai-safety-commitments-ai-seoul-summit-2024 (2024; updated 7 February 2025) (accessed 3 July 2026).
15. Slattery, P. et al. The AI Risk Repository: a comprehensive meta-review, database, and taxonomy of risks from artificial intelligence. Preprint at https://arxiv.org/abs/2408.12622 (2024; database updated 2026) (accessed 3 July 2026).
16. Bengio, Y. et al. *International AI Safety Report 2026*. DSIT 2026/001 (UK Department for Science, Innovation and Technology, 2026); preprint at https://arxiv.org/abs/2602.21012 (accessed 3 July 2026).
17. Maynard, A. D. & Garbee, E. Responsible innovation in a culture of entrepreneurship: a US perspective. In *International Handbook on Responsible Innovation: A Global Resource* (eds von Schomberg, R. & Hankins, J.) 488–502 (Edward Elgar, 2019).
18. Maynard, A. D., Oye, K. A., Scragg, M., Tripp, T. & Wolf, S. M. Successfully bridging innovation and application: exploring the utility of a risk innovation approach in the NSF Engineering Research Center for Advanced Biopreservation Technologies (ATP-Bio). *J. Law Med. Ethics* **52**, 553–569 (2024); https://doi.org/10.1017/jme.2024.126
19. Risk Innovation Nexus. Tools and resources for identifying and navigating orphan risks, including the *Risk Innovation Planner* and *Culminating Report* (Arizona State University, School for the Future of Innovation in Society & J. Orin Edson Entrepreneurship + Innovation Institute, 2017–2020); https://riskinnovation.org (accessed 3 July 2026).
20. Douglas, M. & Wildavsky, A. *Risk and Culture: An Essay on the Selection of Technological and Environmental Dangers* (Univ. California Press, 1982); https://www.jstor.org/stable/10.1525/j.ctt7zw3mr
21. Porter, T. M. *Trust in Numbers: The Pursuit of Objectivity in Science and Public Life* (Princeton Univ. Press, 1995); https://www.jstor.org/stable/j.ctt7sp8x
22. Karnofsky, H. Responsible Scaling Policy v3. *LessWrong* https://www.lesswrong.com/posts/HzKuzrKfaDJvQqmjh/responsible-scaling-policy-v3 (24 February 2026) (accessed 3 July 2026).
23. Koessler, L., Schuett, J. & Anderljung, M. Risk thresholds for frontier AI. Preprint at https://arxiv.org/abs/2406.14713 (2024) (accessed 3 July 2026).
24. Hackenburg, K. et al. The levers of political persuasion with conversational artificial intelligence. *Science* **390**, eaea3884 (2025); https://doi.org/10.1126/science.aea3884
25. Kasirzadeh, A. Two types of AI existential risk: decisive and accumulative. *Philos. Stud.* **182**, 1975–2003 (2025); https://doi.org/10.1007/s11098-025-02301-3
26. Power, M. *The Audit Society: Rituals of Verification* (Oxford Univ. Press, 1997).
27. Stelling, L. et al. Evaluating AI providers' frontier safety frameworks. Preprint at https://arxiv.org/abs/2512.01166 (2025) (accessed 3 July 2026).
28. Vaughan, D. *The Challenger Launch Decision: Risky Technology, Culture, and Deviance at NASA* (Univ. Chicago Press, 1996).
29. Abbott, J. & Maynard, A. *AI and the Art of Being Human: A Practical Guide to Thriving with AI While Rediscovering Yourself in the Process* (Waymark Works Publishing, 2025).
30. International Organization for Standardization. *ISO 31000:2018 Risk Management — Guidelines* (ISO, 2018); https://www.iso.org/standard/65694.html; *ISO/IEC 23894:2023 Information Technology — Artificial*

*Intelligence — Guidance on Risk Management* (ISO, 2023); https://www.iso.org/standard/77304.html (accessed 3 July 2026).

31. Kasperson, R. E. et al. The social amplification of risk: a conceptual framework. *Risk Anal.* **8**, 177–187 (1988); https://doi.org/10.1111/j.1539-6924.1988.tb01168.x
32. Stilgoe, J., Owen, R. & Macnaghten, P. Developing a framework for responsible innovation. *Res. Policy* **42**, 1568–1580 (2013); https://doi.org/10.1016/j.respol.2013.05.008
33. Maynard, A. D. Why we need risk innovation. *Nat. Nanotechnol.* **10**, 730–731 (2015); https://doi.org/10.1038/nnano.2015.196
34. Heaven, W. D. Why Meta's latest large language model survived only three days online. *MIT Technology Review* (18 November 2022); https://www.technologyreview.com/2022/11/18/1063487/meta-large-language-model-ai-only-survived-three-days-gpt-3-science/ (accessed 3 July 2026).
35. Mickle, T., Metz, C., Isaac, M. & Weise, K. Inside OpenAI's crisis over the future of artificial intelligence. *The New York Times* (9 December 2023); https://www.nytimes.com/2023/12/09/technology/openai-altman-inside-crisis.html (accessed 3 July 2026).
36. Leike, J. Posts on X (17 May 2024); https://x.com/janleike/status/1791498174659715494; archived at https://web.archive.org/web/20260609000957/https://x.com/janleike/status/1791498174659715494 (accessed 3 July 2026).
37. Raine v. OpenAI, Inc., No. CGC-25-628528 (Super. Ct. Cal., Cty. of San Francisco, filed 26 August 2025); complaint archived at https://web.archive.org/web/20260601012712/https://www.courthousenews.com/wp-content/uploads/2025/08/raine-vs-openai-et-al-complaint.pdf (accessed 3 July 2026).
38. OpenAI. *OpenAI Charter* (OpenAI, 2018); https://openai.com/charter (accessed 3 July 2026).
39. Maynard, A. D. The AI cognitive Trojan horse: how large language models may bypass human epistemic vigilance. Preprint at https://arxiv.org/abs/2601.07085 (2026) (accessed 3 July 2026).
40. Shaw, S. D. & Nave, G. Thinking — fast, slow, and artificial: how AI is reshaping human reasoning and the rise of cognitive surrender. Preprint at https://doi.org/10.31234/osf.io/yk25n_v1 (2026).